\documentclass[useAMS,usenatbib]{mn2e}
\usepackage{graphicx}

\title[The RGB Tip and Bump of the Leo~II dSph]
{The Red Giant Branch Tip and Bump of the Leo~II dwarf spheroidal galaxy
 \author[M. Bellazzini et al.]{M. Bellazzini$^{1}$, 
N. Gennari$^{2}$, F.R. Ferraro$^{2}$
\thanks{E-mail: bellazzini, ferraro@bo.astro.it}\\
$^{1}$INAF - Osservatorio Astronomico di Bologna, via Ranzani 1, 40127, Bologna,
Italy\\
$^{2}$Universit\`a di Bologna - Dipartimento di Astronomia, via Ranzani 1, 40127, Bologna,
Italy}
}
\begin{document}

\date{\today}

\pagerange{\pageref{firstpage}--\pageref{lastpage}} \pubyear{2003}

\maketitle

\label{firstpage}

\begin{abstract}
We present V and I photometry of a $9.4\arcmin \times 9.4\arcmin$ field 
centered on the dwarf spheroidal galaxy Leo~II.  The Tip of the Red Giant
Branch is identified at $I^{TRGB}=17.83 \pm 0.03$ and adopting $\langle
[M/H]\rangle = -1.53 \pm 0.2$ from the comparison of RGB stars with Galactic
templates, we obtain a distance modulus $(m-M)_0=21.84 \pm 0.13$, corresponding
to a distance $D=233 \pm 15$ Kpc. Two significant bumps have been detected in
the Luminosity Function of the Red Giant Branch. The fainter bump (B1, at
$V=21.76\pm 0.05$) is the RGB bump of the dominant stellar population while the
actual nature of the brightest one  (B2, at $V=21.35\pm 0.05$) cannot be firmly
assessed on the basis of the available data, it can be due to the Asymptotic
Giant Branch Clump of the  main population or it may be a secondary RGB bump. 
The luminosity of the main RGB bump (B1) suggests that the majority of  RGB
stars in Leo~II belongs to a population that is $\ga 4$ gyr younger than the
classical Galactic globular clusters. The stars belonging to the He-burning Red
Clump are shown to be significantly more centrally concentrated than RR Lyrae
and Blue Horizontal Branch stars, probing the existence of an age/metallicity
radial gradient in this remote dwarf spheroidal. \end{abstract}


\section{Introduction}

Leo~II \citep{hw} is one of the most distant  dwarf spheroidal (dSph)
satellites of  the Milky Way \cite[$D\simeq 205$ Kpc, according to][]{mateo}.  
From deep HST photometry \citet{ken} determined a mean age of $9 \pm 1$ Gyr for
the main stellar population of this galaxy, with a Star Formation History (SFH)
started $\sim 14$ Gyr ago and lasted for $\sim 7$ Gyr.  According to these
authors, after this epoch the SF in Leo~II has been very low and, at most,
sporadic \cite[see also][that reached similar conclusions following a different
path]{bosler}. The presence of a significant  fraction of old stars (e.g., age
$\ga 10$ Gyr) in this galaxy is witnessed by a conspicuous population of RR
Lyrae variables  \cite[see][and references therein]{siegel} and Blue Horizontal
Branch (BHB) stars \citep{demers,ken}. 

The observed color of the Red Giant Branch (RGB) is typical of a metal poor
population, and most of the existing photometric studies 
\citep{demharris,lee_leo,demers} agree in deriving an average metallicity
$[Fe/H]\simeq -1.9$ in the \citet[][hereafter ZW]{zw} metallicity scale, with a
metallicity dispersion of $\sim 0.3$ dex, while \citet{ken} obtained
$[Fe/H]\simeq -1.6$.  During a recent low-resolution spectroscopic survey,
\cite{bosler} determined  the metallicity of 41 RGB stars in Leo~II, in the
\citet[][hereafter CG]{cg} metallicity scale. The metallicity of the stars
studied by \cite{bosler} ranges from $[Fe/H]_{CG}=-2.32$ to
$[Fe/H]_{CG}=-1.26$; the average metallicity is $\langle [Fe/H]_{CG}\rangle =
-1.57$, in good agreement with the photometric estimates described above, once
the differences in the metallicity scales are taken into account. A global
mass-to-light ratio $M/L\simeq 11.1\pm 3.8$  \citep{vogt} suggests that the
Dark Matter (DM) is the main contributor to the total mass of the galaxy, as in
most of the other dSphs of the Local Group. 

As a part of a large programme aimed at obtaining homogeneous distances for
most of the galaxies of the Local Group \cite[see][]{draco,m33,leo1,sgrtip}, we
present here the results of the V,I photometry (reaching $V\simeq 23$) of a 
$9.4\arcmin \times 9.4\arcmin$ field centered on Leo~II. We provide a new
estimate of the distance to this  galaxy using the Tip of the Red Giant Branch
(TRGB) as a standard candle \citep[see][for details and references about the
method] {lfm93,smf96,mf98,scw,tip1,tip2}, and we study the main properties of
the evolved stellar populations of the galaxy. In the present analysis we
consider a field of view significantly larger than in previous studies.
Adopting a King profile for the light distribution of Leo~II, with core radius
$r_c=2.9\arcmin$ and tidal radius $r_t=8.7\arcmin$ \cite[according
to][]{mateo}, we estimate that our field cover a fraction of the area of 
the galaxy ($\simeq
37$\%) that includes more than 87\% of the total integrated light emitted by
Leo~II. The corresponding figures (fraction of the area/fraction of the light)
for previous studies are :  20\%/70\% \citep{demers}, 16.6\%/63\%
\citep{lee_leo}, 11\%/50\% \citep{demharris}, 2\%/15\% \citep{ken}. The larger
radial range explored by the present study may be at the origin of some
differences in the estimates of the average properties  of the galaxy  (as, for
example the mean metallicity) with respect to  previous analyses, due to radial
population gradients  (see Sect.~3.2 and 5., below).

The plan of the paper is the following: in Sect.~2 we describe the
observational material, the data reduction process, the artificial stars
experiments, we briefly discuss the Color Magnitude Diagrams and we present the
light-curve of a newly identified variable; in Sect.~3 we report on the
detection of the TRGB, and on our estimate of the distance modulus. In Sect.~4
we discuss the two  bumps we identify in the Luminosity Function of the RGB and
in Sect.~5 we present and discuss the radial distribution of HB stars, showing
the  presence of a population gradient.  Finally, the main results  are
summarized in Sect.~6.

\section{Observations, Data Reduction and Color Magnitude Diagrams}

The  data were obtained  at the  3.52 $m$  Italian telescope  TNG  (Telescopio 
Nazionale  Galileo - Roque  de  los  Muchachos, La  Palma, Canary Islands,
Spain),  using DoLoRes, a focal reducer imager/spectrograph   equipped with a
$2048 \times 2048$ pixels CCD array.  The pixel scale is $0.275$  arcsec/px.
The observations were carried out during  three nights (March 19, 20 and 21,
2001), under average seeing conditions  ($FWHM\simeq  1.0\arcsec -
1.4\arcsec$). The data have been acquired during the same observational run
already described in \citet{draco,leo1}: any further detail may be found in 
those papers.

We acquired five exposures in I (two with $t_{exp}=600$ s, and three with
$t_{exp}=300$ s), and eight exposures in V (two with $t_{exp}=600$ s, and  six
with $t_{exp}=300$ s), centered on the center of Leo~II.  All the  raw  images
were  corrected  for bias  and  flat  field, and  the overscan  region  was 
trimmed  using standard  IRAF\footnote{IRAF  is distributed by  the National 
Optical Astronomy Observatory,  which is operated by the Association of
Universities for Research in Astronomy, Inc.,   under  cooperative   agreement 
with   the   National  Science Foundation.} procedures. Each set of images  was
registered,  flux-normalized and combined into one single {\em master frame}
(per filter) using the  tasks {\em interp.csh} and {\em ref.csh} of the 
ISIS-2.1 package \citep{alard}. ISIS is able to combine images into a master 
frame having the seeing of the best image in the set, without any loss of flux
\citep{alard,alard1}\footnote{See also the on-line tutorial 
http://www2.iap.fr/users/alard/tut.html}.

The PSF-fitting procedure was performed  independently on each V and I master
image, using a version of DoPhot \citep{doph} modified by   P. Montegriffo at
the Bologna  Observatory to  read  images in  double precision  format. The
frames were searched for sources adopting a 5-$\sigma$ threshold, and the
spatial variations of the PSF were modeled with a quadratic polynomial. The
adopted threshold corresponds to the limiting magnitudes  $I\simeq 22.2$ and
$V\simeq 23.2$. A  final  catalogue listing  the instrumental  V,I magnitudes 
for all  the  stars in  each  field has  been obtained  by cross-correlating
the V and  I catalogues. Only the sources classified as  stars  by  the  code 
have been  retained. Aperture corrections have been determined on a sample of
bright and isolated stars in each of the master frames and applied to the
catalogues.

The transformation to the standard Johnson-Cousins photometric system has been
achieved using the calibrating relation obtained and described in
\citet{draco}. The absolute calibration has been checked against independent
photometries \citep{stet,momany} and it has been found to be accurate at the
$\pm 0.02 $ mag level \cite[see][]{draco}. The astrometric solution to
transform the original reference frame (X,Y in pixels) to equatorial
coordinates at the 2000.0 Equinox have been obtained from 58 stars in common
with the GSC2.2 catalogue\footnote{see
http://www-gsss.stsci.edu/gsc/gsc2/GSC2home.htm}, using a dedicated package
developed by P. Montegriffo at the Bologna Observatory. The root-mean-square
residuals of the transformation are $< 0.2\arcsec $ in both RA and Dec. 

We cross-correlated our catalog with the Near Infrared Point Source Catalogue 
of 2MASS \citep{cutri} and we found 18 sources in common, among the brightest
and reddest stars of our dataset. The root-mean-square between our astrometric
solution and the astrometry from 2MASS for the 18 stars in common is $<
0.15\arcsec$ in both RA and Dec. We identified also 5 carbon stars from
\citet{carb} and 111 RR Lyrae and 2 anomalous cepheids from the catalog of
\citet{siegel}. The final catalogue is available in electronic form from the
CDS.

$E(B-V)=0.02\pm 0.01$, according to \citet{mateo} \cite[and in agreement
with][]{cobe} and  $A_I=1.76E(B-V)$, according to \citet{dean}, are adopted
throughout the following analysis.

\begin{figure} 
\includegraphics[width=84mm]{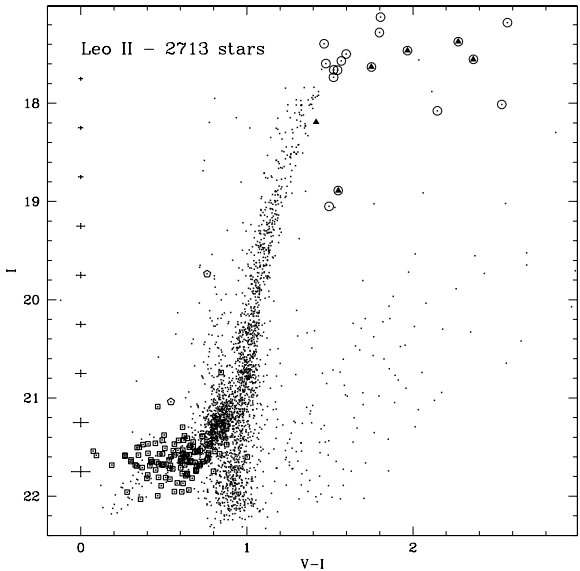}  
\caption{I,V-I CMD of Leo~II. The
characteristic photometric uncertainties are plotted in the two panels as
errorbars aligned at $V-I=0.0$. Large open circles mark stars in common with
the Point Source Catalogue of 2MASS. Filled triangles are carbon stars we have
identified from \citet{carb}. Pentagons are anomalous Cepheid (V53 and V171)
and open squares are RR Lyrae identified from \citet{siegel}.} 
\end{figure}

\subsection{Color-Magnitude Diagram}

The Color Magnitude Diagram (CMD) obtained from our catalogue is displayed in
Fig.~1. The morphology of the CMD is fully consistent with those presented and
described in previous studies that attained sufficiently deep and accurate
photometry to unveil the morphology of the entire Horizontal Branch
\citep{demers,ken,momany}.

The CMD is dominated by the steep RGB of Leo~II extending from the limiting
magnitude to $I\simeq 17.8$. A handful of bright AGB stars, most of which are
also detected in the infrared by 2MASS (marked by open circles in Fig.~1),  are
also present at $I<18.0$ and $V-I\ge 1.4$. Most of the identified carbon stars
are included in this family. The star detected by 2MASS at $I\simeq 19.0$ and
$V-I\simeq 1.4$ may be an unknown carbon star but a check of the membership
should be performed to obtain a firm classification. This star has $J-K\simeq
1.2$, similar to that of the recognized carbon stars.

The majority of He-burning stars are clustered in a Red Clump around  $I\simeq
21.3$ and $V-I\simeq 0.9$, but a significant number of RR Lyrae is also present
and a sparse tail of BHB stars is clearly visible at $V-I\le 0.4$. The sparse
plume of stars between $I\sim 21$ and $I\sim 19.5$,  at $V-I\simeq 0.8$
probably hosts the high-mass tail of He-burning stars of  the galaxy. One of
the two Anomalous Cepheid contained in our catalogue lies at the tip of this
feature, in agreement with the above hypothesis. According to the Galactic
model by \citet{robin} the total number of foreground Milky Way stars sampled
by our field in the color and magnitude ranges displayed in Fig.~1 is $\simeq
21$, with only three foreground stars falling into the  region of the CMD
populated by the above described plume.

\subsubsection{A new RR Lyrae variable}

The package we used to produce the above described master frames
\cite[ISIS][]{alard} is aimed at the search of variable stars by mean of a very
efficient image subtraction technique  \cite[see, for
example][]{sam,kal,gisella}. Just to verify the performance of the code in
difficult conditions (i.e., very sparse time series and typical target just
$\sim 1$ mag brighter than the limiting magnitude of the photometry) we applied
the whole image subtraction process to the eight V images from which we
obtained our master frame. We identified a list of 14 bona-fide variables with
detected variations larger than 0.1 mag in the considered  lapse of time.
Thirteen of them were variables included in the catalogue of \citet{siegel}, 12
ab type RR Lyrae and 1 RRc. The fourteenth identified variable is not included
in Siegel et al.'s catalogue but the partial light curve we obtained (Fig.~2,
upper panel) indicates that it is a real variable star. The large amplitude
($>0.5$ mag) and its position in the CMD (shown in the lower panel of Fig.~2)
suggest a preliminary classification as a RRab. Following the numbering by
\citet{siegel} we dub this variable V173. Its coordinates are  $(RA_{2000}~;~
Dec_{2000})=(11^h 13^m 23.74^s ~;~ 22\degr 07\arcmin 21.7\arcsec)$.

\begin{figure} 
\includegraphics[width=84mm]{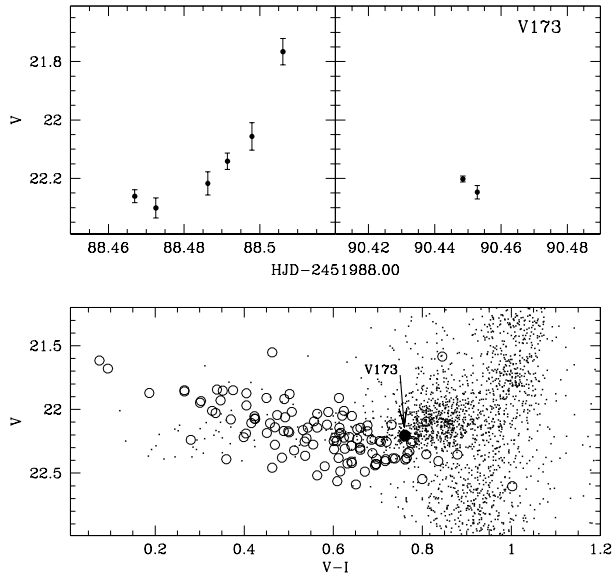} 
\caption{Upper panels: Temporal light
variations of the newly identified candidate RR Lyrae star V173. The position
of V173 is marked by a filled circle in the CMD shown in the lower panel. Open
circles are the RR Lyrae identified from \citet{siegel}.} 
\end{figure}

\subsection{Artificial Stars Experiments}

To quantify the effects of the data reduction process on our photometry we
performed a set of artificial stars experiments.  We followed exactly the
procedure described in \citet{draco} and we refer the interested reader to that
paper for any detail. The artificial stars were extracted from a LF similar to
the observed one, with the additional requirement that they must lie on the
average ridge line representing the observed RGB. We limited the artificial
stars experiments to RGB stars with $I\le 22.0$.  The stars were added ($\sim
100$ at a time to avoid any spurious modification of the actual crowding
conditions) to the master frames and the whole process of data reduction was
repeated at any run. A total of $\ga 10^4$ artificial stars have been added and
processed on the V and I master frames.

\begin{figure} 
\includegraphics[width=84mm]{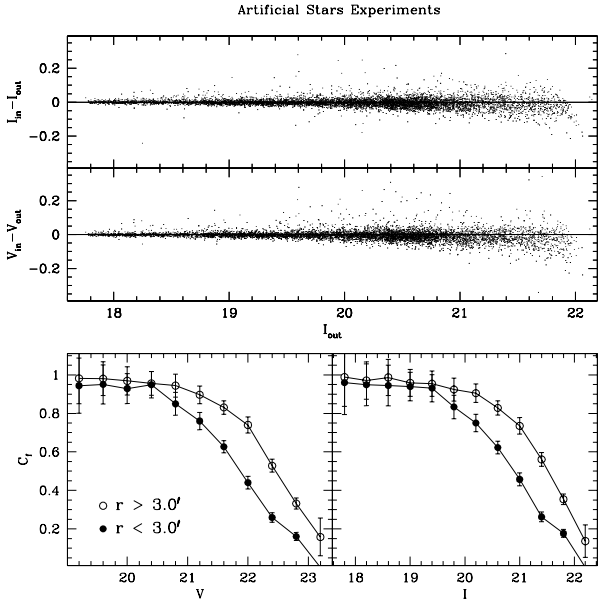}  
\caption{Results of the
artificial stars experiments. Upper panels: differences between input and
output magnitudes as a function of output magnitude $I_{out}$. Lower panels:
the completeness factor \citep[$C_f$, see][]{draco,leo1} as a function of V and
I magnitudes. The effects of the variation of the completeness with the
distance from the center of the galaxy are shown by the different behaviour of
$C_f(V)$ and $C_f(I)$ in the region within $3\arcmin$ from the center (filled
circles) and outside $3\arcmin$ from the center (empty circles).  }
\end{figure}

The difference between input and output magnitudes, shown in the upper panels
of Fig.~3, confirms that the photometric errors are quite small and the degree
of blending cannot affect our analyses. The total completeness of the sample 
(Fig.~3, lower panels) is larger than 80\%  for $I(V)\la 20.0(21.0)$ and  drops
below 50\% for $I(V)\ga 21.0(22.0)$. The increasing of stellar crowding toward
the center of the galaxy produces a significant radial variation of the
completeness. The two curves plotted in the lower panels of Fig.~3 show the
behaviour of $C_f(V)$ and $C_f(I)$ in the regions within $3\arcmin$ from the
center (filled circles) and outside $3\arcmin$ from the center (empty circles).
The spatial variation of the completeness cannot affect any of the results and
analyses reported in the following, with the only possible exception of the
radial population gradient presented in Sect.~5, below (see the same Sect.~5
for discussion).

\section{The distance to Leo~II}

\subsection{Detection of the TRGB} 

The use of Tip of the Red Giant Branch (TRGB) as a standard candle is now a
mature and widely used technique to estimate the distance to galaxies of any
morphological type  \cite[see][for a detailed description of the method, recent
reviews and applications]{lfm93,mf95,mf98,walk}. The underlying physics is well
understood \citep{mf98,scw} and the observational procedure is operationally
well defined \citep{mf95}. The key observable is the sharp cut-off occurring at
the bright end of the RGB Luminosity Function (LF) that can be easily detected
with the application of an edge-detector filter \citep[Sobel
filter,][]{mf95,smf96} or by other (generally parametric) techniques \cite[see,
for example][]{mendez,mcc}. The necessary condition for a safe application of
the technique is that the observed RGB LF should be well populated, with more
than $\sim 100$  stars within 1 mag from the TRGB \citep{mf95,draco}.  The
present sample is at the limit of this requirement, having  $N_{\star}=106$ 
stars within one magnitude from the detected TRGB  (see below). In Sect.~3.3 we
will provide additional evidence supporting the robustness of our TRGB
detection.

The detection of the TRGB is displayed in Fig.~4. The cut-off of the RGB LF is
clearly evident and it is easily detected by the Sobel's filter at
$I^{TRGB}=17.83 \pm 0.03$, where the reported uncertainty is the Half Width  at
Half Maximum of the peak of the filter response.

\begin{figure} 
\includegraphics[width=84mm]{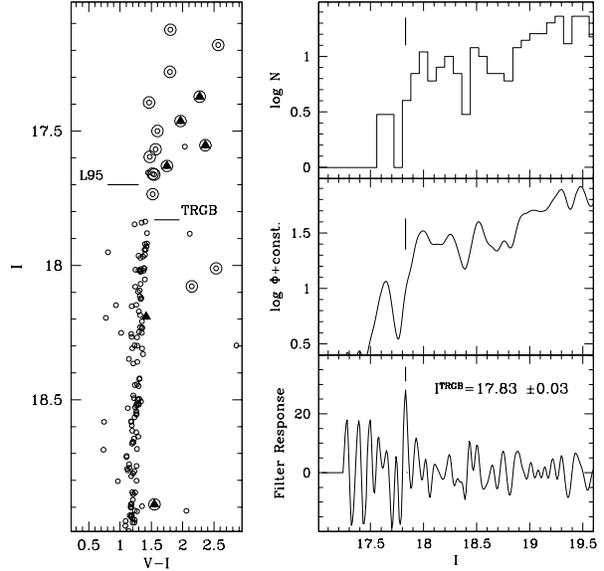} 
\caption{TRGB detection in Leo~II. 
The logarithmic LF of the upper portion of the RGB is shown as an ordinary 
histogram (upper right panel) and as a generalized histogram  \citep[middle
right panel, see][]{laird}. The lower right panel shows the Sobel's filter
response to the LF.  In all panels the thick line marks the position of the
TRGB. Left panel: the TRGB level found by us (TRGB) and by \citet{lee_leo}
(L95) are reported on the CMD. The symbols are the same as in Fig.~1. }
\end{figure}

The only previous published estimate of $I^{TRGB}$ for Leo~II is from
\citep[][hereafter L95]{lee_leo} who found $I^{TRGB}=17.7\pm 0.2$.  While
formally consistent with our measure, within the errors, the CMD shown in the
left panel of Fig.~4 suggests that the TRGB level identified by L95 is too
bright, probably due to the influence of the AGB population on his LF. In this
regard, it has to be considered that our field cover an area that is double
with respect to L95 and that our sample collects 30\% more stars than that of
L95, down to $I=22.0$.

\subsection{Metallicity}

In Fig.~5 the observed RGB of  Leo II is compared with the template ridge lines
of the globular clusters NGC~6341, NGC~6205 and NGC~288,  taken from the set
adopted by \citet{m31}.  It is immediately clear that the bulk of Leo~II RGB
stars is enclosed  within the ridge lines of NGC~6341 ($[Fe/H]_{CG}=-2.16$;
$[Fe/H]_{ZW}=-2.24$; $[M/H]=-1.95$) and of  NGC~6205 ($[Fe/H]_{CG}=-1.39$;
$[Fe/H]_{ZW}=-1.65$; $[M/H]=-1.18$)\footnote{[M/H] is the {\em global
metallicity}, a parameter that includes the contribution of Iron  and of the
$\alpha$-elements (O, Mg, Ti, Si, etc.),  hence it is a more suitable indicator
of the global metal content to relate with the  observed properties of stars
and stellar populations and for comparisons with theoretical models
\cite[see][for discussion and references]{scs93,f99,tip2,leo1}.}. 

We derived individual photometric estimates of the metallicity in the 
different scales  for 165 RGB stars having $M_I< -2.5$ by interpolating on the
grid of  templates ridge lines, as done in \citet{draco,m31} and \citet{m33}. 
From the obtained distributions we derived robust estimates of the average
metallicity and of the standard deviation. The median and the average of all
the considered distributions differ by just $\la 0.03$ dex, indicating that the
average  values are not driven by few outlier points but are truly
representative of the bulk of the distributions. The final estimates of the
mean metallicity and of the distance modulus (see Sect.~3.3, below) have been
obtained by an  iterative process, adopting $M_I^{TRGB}=-4.05$ as initial guess
\cite[see][]{m33}. The process converged immediately to the final values
because the dependence of  $M_I^{TRGB}$ on metallicity is essentially null in
the range of metallicity around $[M/H]\sim -1.5$. In the following of this
section we shortly compare our average metallicities with previous estimates
found in the literature.

\begin{itemize}

\item{} {\em ZW scale:} we obtain $\langle[Fe/H]_{ZW}\rangle=-1.91$ and 
$\sigma=0.22$ dex in excellent agreement with \citet{lee_leo}, \citet{demers} 
and \citet{siegel}. The agreement with \citet{ken} is less satisfying but the
difference of their estimate to all the others (0.3 dex) is not a reason  of
serious concern. Moreover, these authors sampled the very central region of the
galaxy, where the average metallicity of the stars may be intrinsically higher
with respect to outer regions (see Sect.~5, below). 

\item{}{\em CG scale:} we obtain $\langle[Fe/H]_{CG}\rangle=-1.74$ and
$\sigma=0.30$ dex, within 0.2 dex of the spectroscopical estimate by
\citet{bosler}. Our metallicity distribution  is also broadly similar to that
obtained by these authors (for instance, the range is $-2.4\le [Fe/H]_{CG}\le
-1.2$ while Bolster et al. find $-2.32\le [Fe/H]_{CG}\le -1.26$), suggesting
that our photometric metallicities are quite reliable in the present case.

\item{}{\em global metallicity scale:} $\langle[M/H]\rangle=-1.53$ and
$\sigma=0.30$ dex. Taking into account the  contribution of photometric errors
to the derived metallicity dispersion as done in \citet{draco,m33}, we estimate
that the $1-\sigma$ intrinsical dispersion is $\simeq 0.25$ dex, in good
agreement with \citet{lee_leo,demers,ken,bosler}.

\end{itemize} 

\begin{figure}  
\includegraphics[width=84mm]{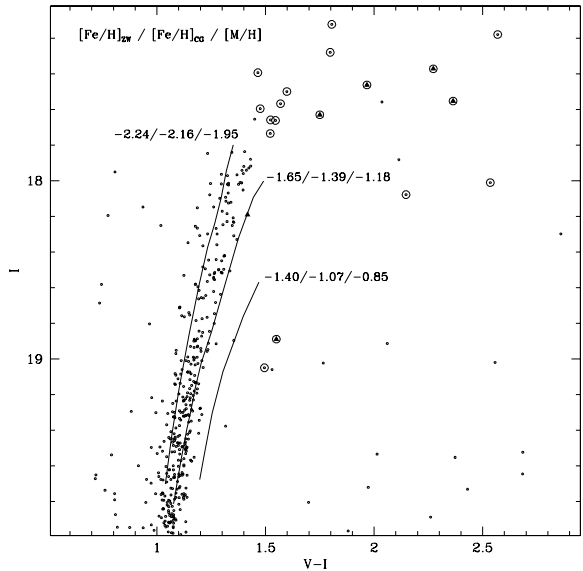}   
\caption{Comparison of
the observed RGB of Leo~II (assuming $(m-M)_0=21.84$ and $E(B-V)=0.02$) with 
the ridge lines of the templates adopted by \citep{m31}. The plotted ridge
lines are from NGC~6341, NGC~6205 and NGC~288, from left to right,
respectively. The label reports the metallicity of the templates in the ZW, CG
and global metallicity scales. The symbols are the same as in Fig.~1.}
\end{figure}

\subsection{The distance modulus of Leo~I} 

To obtain the distance modulus of Leo~II we adopt the calibration of
$M^{TRGB}_I$ as a function of the global metallicity ($[M/H]$) recently
provided by \citet{tip2} \begin{equation}
M^{TRGB}_I=0.258[M/H]^2+0.676[M/H]-3.629 ~~~\pm 0.12. \end{equation}  At the
mean metallicity  $[M/H]=-1.53$  (as derived in Sect.~3.2, above), the
resulting distance modulus  is $(m-M)_0=21.84$.

This estimate is affected by the combination of uncertainties coming from
different sources, i.e. : the estimate of apparent magnitude of the TRGB
($\sigma=0.03$ mag, plus the $\sigma=0.02$ mag uncertainty on the zero-point of
the absolute photometric calibration),  the calibrating relation
\cite[$\sigma=0.12$ mag, see][]{tip2}, the reddening ($\sigma=0.01$ mag), and
the assumed global metallicity ($\sigma=0.3$ dex). Properly propagating all
these  uncertainties we finally obtain $(m-M)_0=21.84\pm 0.13$,  corresponding
to a heliocentric distance $D=233\pm 15$ Kpc. 

This result is consistent with all previous estimates within the 
uncertainties  quoted by the various authors, typically $\pm 0.2$ mag, but is
outside of their formal range  $21.55\le (m-M)_0\le 21.78$. However it has to
be considered that (a) all previously determined distance moduli are tied to
different  RR Lyrae-based distance scales and (b) the uncertainty in the
absolute  magnitude of the adopted standard candle was taken as null in all
cases.  Hence (a) part of the difference may be due to systematic differences
between  the adopted distance scales and (b) the true $1-\sigma$ uncertainties
of previous  estimates should be of order $\sim \pm 0.25/0.3$ mag, to compare
with our  $\pm 0.13$ which includes all the possible sources of errors. 

\begin{figure}  
\includegraphics[width=84mm]{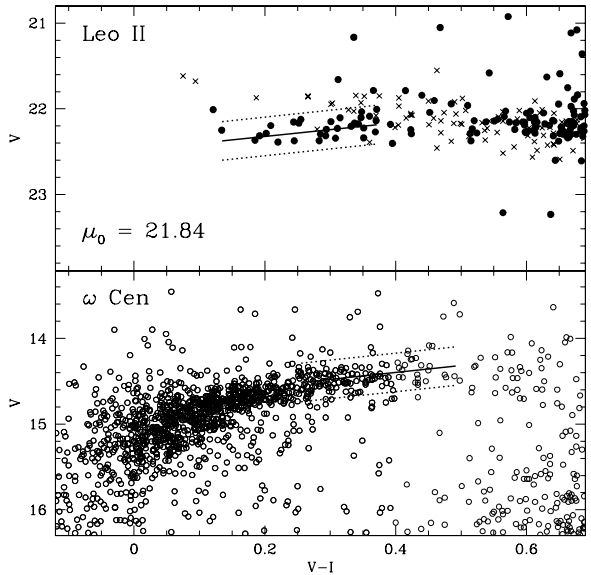}   
\caption{Upper panel:
the HB of Leo~II. The thick continuous line is the ridge line of the BHB stars,
while the dotted lines enclose the whole  BHB distribution. Lower panel: the
Leo~II BHB ridge lines are compared with the observed BHB stars of $\omega$ Cen
\citep[data from][]{panc} after correction for the differences in reddening and
distance modulus between the two stellar systems. The adoption of the distance
modulus derived from the TRGB provides an excellent match of the BHB
sequences.} 
\end{figure}

As said in Sect.~3.1, the number of bright RGB stars of Leo~II contained in our
sample is just above the limit for a safe detection of the TRGB. To verify the
robustness of the above derived distance modulus to sampling effects we compare
the luminosity of the BHB of Leo~II with that of the cluster that provide the
fundamental zero-point of our calibration of the TRGB method  \cite[e.g.
$\omega$ Centauri, see][]{tip1,tip2}. For a favorable circumstance the
metallicity of the main population of $\omega$ Cen ($[M/H]\simeq -1.4$, see
\citet{sollima}) is very similar to the average metallicity of Leo~II. In the
upper panel of Fig.~6 we show the ridge line of the mean level of the BHB of
Leo~II (continuous line) as well a the lines enclosing the whole BHB
distribution (dotted lines). In the lower panel of Fig.~6 these ridge lines 
are compared to the observed BHB of $\omega$ Cen \cite[from][]{panc} adopting
$(m-M)^{Leo~II}_0=21.84$ and $(m-M)^{\omega Cen}_0=13.70$, $E(B-V)^{\omega
Cen}=0.11$,  according to \citet{tip2}. It can be readily appreciated that the
TRGB distance modulus carries the HBs of Leo~II and $\omega$ Cen essentially to
a perfect match, providing further support to the reliability of our distance
estimate.

\section{The RGB bump(s)}

The RGB bump is the effect of a well known phase of the RGB evolution of low
mass stars \cite[see][and references therein]{iben,ffp}. When the H-burning
shell of a RGB star encounters the chemical discontinuity left behind by the
maximum penetration of the convective envelope, the luminosity of the star has
a slight drop. When the shell adapts at the new environment the luminosity 
grows again, but at a different pace than before. As a result the stars pile-up
at this stage, producing a {\em bump} in the RGB LF and the slope of the LF
changes above the bump, because of the change in the evolutionary rate. The RGB
bump has been extensively studied in globular clusters  \cite[see][- hereafter
F99 - and references therein]{ffp,z99,f99},  but it has been detected in dwarf
galaxies only in recent times \citep{sculp,sex,draco,sgr,lee_sex}. \citet{sgr}
and \citet{bump} provided direct evidence that significant  constraints on the
age and metal content of dSph galaxies can be obtained  from the study of this
observational feature. 

\citet{sculp} and \citet{sex} interpreted the detection of two bumps on the RGB
LF as indication of the presence of two distinct populations in Sculptor and
Sextans, respectively. A preliminary detection of a double RGB bump in Leo~II,
from the same dataset studied here, has been presented in \citet{bump}.
\citet{lee_sex} argued against this interpretation, proposing that the
brightest of the two bumps found in these galaxies it is due to the AGB clump
instead \citep{carma}. 

In the attempt of minimizing the effects of any possible AGB Clump on the RGB
LF we carefully selected our sample of RGB stars. The adopted selection is
shown in the upper-right panel of Fig.~7. Note that the sparse cluster of stars
located at $V\simeq 21.3$ and $V-I\simeq 0.9$ that may be identified with the
AGB clump of the main population of Leo~II has been excluded from the
selection, as well as the bluest stars of the whole sequence, which may
presumably have the highest degree of contamination from AGB stars. In spite of
that, the right-panels of Fig.~7 show that two significant bumps are detected
on the RGB of Leo~II. The main bump (B1) is at $V^{bump}=21.76\pm 0.05$, the
less pronounced one (B2) is at $V^{bump}=21.35\pm 0.05$.   While B1 can be
straightforwardly interpreted as the RGB bump of the  dominant stellar
population of Leo~II, the nature of B2 is more uncertain.  We combined Eq.~7
and Eq.~6.6 of F99  (obtained from well studied galactic globulars) to derive a
relation for the difference in V magnitude of the AGB clump and the RGB Bump
($\Delta V^{AGB}_{bump}$) as a function of the global metallicity:
\begin{equation} \Delta V^{AGB}_{bump}=-0.360[M/H]^2-1.772[M/H]-2.283 ~~~\pm
0.09. \end{equation} The difference in magnitude between B2 and B1 ($\Delta
V_{B2-B1}=-0.41\pm0.07$)  is in excellent agreement with the predictions of 
Eq.~2 for $[M/H]\simeq -1.5$, i.e. $\Delta V^{AGB}_{bump}=-0.43\pm
0.09$\footnote{At odds with the results of the preliminary analysis shown in
\citet{bump}. The difference is entirely due to the fact that in that analysis
we assumed  $[M/H]\simeq -1.7$, instead of $[M/H]\simeq -1.5$, as adopted
here.}. This result seems to favor the interpretation of B2 as a feature
associated with the AGB clump of the same population that generated the RGB
bump B1. On the other hand, as shown in Fig.~7, our RGB selection should have
removed most of the AGB clump stars from the considered sample.  Hence the
possibility that B2 is a genuine secondary RGB Bump cannot be excluded at the
present stage. Given this uncertain interpretation we drop any further
discussion of B2 in the following analysis, which is focused on  B1 instead.
Larger photometric samples and dedicated theoretical modeling of the 
luminosity functions in the AGB phase are needed to achieve a clearer view of
the problem of double bumps in the context of composite stellar populations.

\begin{figure*}  
\includegraphics[width=168mm]{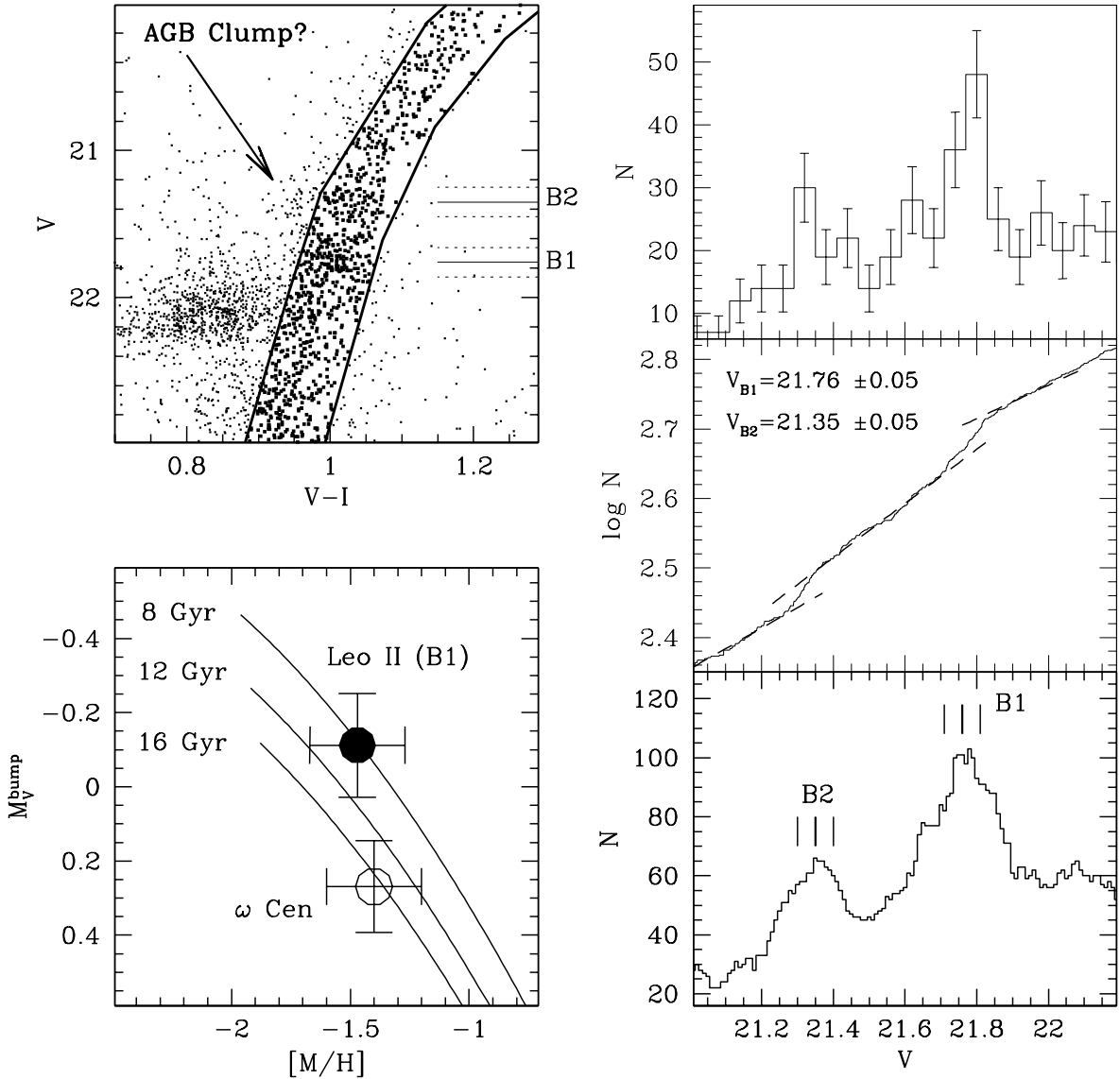}   
\caption{ Upper-left panel: the
selection of RGB stars adopted to derive the LF. The selected RGB stars are the
heavier dots enclosed within the RGB-shaped box. Upper-right panel: the LF as
an ordinary histogram, with error bars. Middle-right panel: the LF as a 
logarithmic cumulative distribution. Lower panel: the LF as a smoothed
histogram with bin width of 0.16 mag and incremental step of 0.01 mag. The
position of the bumps and the associated uncertainty ranges are indicated by
thick and thin vertical segments, respectively. Lower-left panel: The absolute
V magnitude of the RGB bump of Leo~II (B1, filled circle) and of the main
population of $\omega$ Cen (open circle, see S04) are compared with isochrones
of different ages derived from Eq.~3 of F99.}  
\end{figure*}

In the lower-left panel of Fig.~7 we compare the absolute V magnitude of B1 
with three isochrones in the $M_V^{bump}$ vs. $[M/H]$ plane obtained from Eq.~3
of F99 \cite[see][]{sgr,bump}. According to F99, the bulk of the Galactic
globular clusters is comprised between the 16 Gyr and the 12 Gyr isochrones, in
this plane. Hence, the position of B1 indicates that the main population of
Leo~II is more than 4 Gyr younger than the typical Galactic globular, in
excellent agreement with the results of \citet{ken}, obtained from the analysis
of the Main Sequence  Turn Off (MSTO) of Leo~II. The inclusion in the plot of
the main RGB bump of  $\omega$ Cen (see S04) provides again a real case to
compare with:  the difference in $M_V^{bump}$ between this cluster and
Leo~II(B1) is at the $\ge 2-\sigma$ level. It has to be recalled that the age
difference between the Main Population (MP) of Leo~II and classical globular
clusters would be lower than what  read from the lower-left panel of Fig.~7 if
the Helium abundance of  Leo~II - MP is higher than that of the clusters  (see
S04 and references therein).

\section{A population gradient in Leo~II}

Population gradients are a common feature in dSph galaxies  \cite[see][-
hereafter H01 - and references therein]{harbeck}. In all the cases in which a
radial gradient has been found, the sense is invariably that the most
metal-rich/young populations are more centrally concentrated than the
metal-poor/older ones. Since the HB is the evolutionary phase in which low-mass
stars display the maximum sensitivity to metallicity and age \cite[as well as
to a number of other physical parameters, see][]{fp93}, in most cases the
gradients have been detected using HB stars as tracers \citep[H01,][]{draco}.

In Fig.~8 the cumulative radial distributions of Red Clump (RC) stars and of
BHB+RR Ly stars are compared. The plot shows that the usual kind of radial
gradient is present also in Leo~II: RC stars are significantly more centrally
concentrated the BHB+RR Ly stars. A Kolmogorov-Smirnov test quantify the
probability that the two samples are extracted from the same parent
distribution to less than 0.01 \%. While the two considered populations have
essentially  the same average V magnitude (see Fig.~2, above), the BHB+RR Ly
stars are  $\sim 0.5$ mag fainter than the RC in I, hence the BHB+RR Ly sample
may be slightly less complete than the RC one. Is it possible that this fact,
coupled with the radial variation of the completeness discussed in Sect.~2.2,
has originated a spurious difference in the distribution of BHB+RR Ly and RC
stars? To test this hypothesis we plot in Fig.~8 also the distribution of RGB
stars fainter than the RC (Faint RGB).  Note that this sample should be less
complete than the BHB+RR Ly one since its members are fainter in V while having
I magnitudes similar to BHB+RR Ly. Hence, if the detected population gradient
is due to the radial variations of the completeness, the Faint RGB sample
should display a radial distribution even more extended than the BHB+RR Ly.
Fig.~8 shows that this is not the case: the distribution of Faint RGBs is much
more similar to that of RC stars than to the distribution of BHB+RR Ly.
Therefore, the observed difference between the radial distributions of RC and
BHB+RR Ly stars is not due to completeness effects but it traces a real
population gradient.

It is very likely that both age and metallicity variations are at the origin of
the effect. However, since RC stars dominates the HB population of Leo~II it is
reasonable to associate them with the main population of the galaxy, whose mean
age is $\simeq 9$ Gyr, according to \citet{ken}, while RR Lyrae and BHB stars
should be associated with older populations \cite[up to age=14 Gyr][]{ken}. In
this framework the age would be the main driver of the gradient. 

\begin{figure}  
\includegraphics[width=84mm]{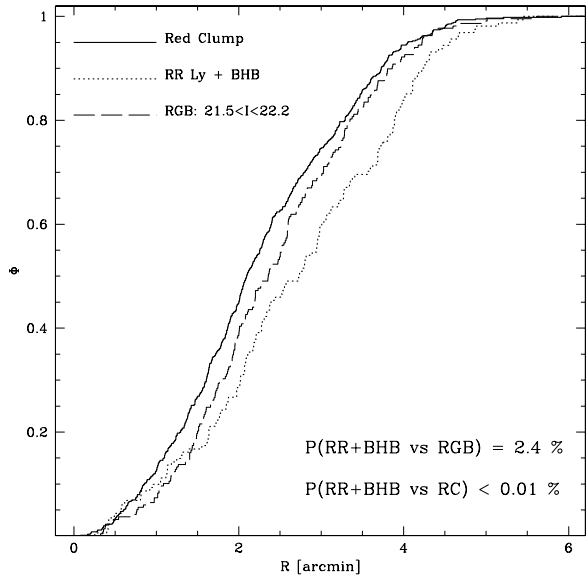}   
\caption{Cumulative
radial distributions of Red Clump stars (continuous line), RR Lyrae and Blue
Horizontal Branch stars (dotted line), and RGB stars fainter than the RC (long
dashed line, adopted as a control sample). The probability that the samples are
extracted from the same parent radial distribution - according to a
Kolmogorov-Smirnov test - is reported in the lower right angle of the plot.} 
\end{figure}

\section{Conclusions}

We have provided a clean and accurate detection of the I magnitude of the TRGB
of Leo~II. Adopting the average metallicity we derived from  the same data by
comparison with templates RGB ridge lines, and the calibration of $M^{TRGB}_I$
as a function of the global metallicity ($[M/H]$) provided by \citet{tip2} we
have obtained a new estimate of the distance modulus of Leo~II, $(m-M)_0=21.84
\pm 0.13$, corresponding to a  distance $D=233\pm 15$ Kpc. The effects of all
the possible sources of uncertainty have been taken into account.  

Two significant bumps have been detected in the LF of the RGB of Leo~II. The
fainter bump (B1) has been identified as the RGB bump of the main population of
the galaxy, while the brighter one (B2) may be due to stars belonging to  the
AGB clump  of the same population or may be a secondary RGB bump, associated
with another population of the galaxy. The luminosity of the main bump (B1)
indicates that the main population of Leo~II is several ($\ga 4$) Gyr younger
than the typical Galactic globular clusters, in good agreement with the
estimates obtained from the photometry of the MSTO by \citet{ken}. This result
suggests that useful indications on the age may be obtained from the RGB bump
also for distant stellar systems whose MSTO is out of the reach of currently
available telescopes \cite[see also][and S04]{sgr,bump}.

A significant population gradient has been detected for the first time in this
galaxy: the BHB and RR Lyrae stars have a more extended radial distribution
with respect to stars in the Red Clump. Age differences are proposed as the
main driver of the observed gradient.

\section*{Acknowledgments} Based on observations made with the Italian
Telescopio Nazionale Galileo (TNG) operated on the island of La Palma by the
Centro Galileo Galilei of the INAF (Istituto Nazionale di Astrofisica)  at the
Spanish Observatorio del Roque de los Muchachos of the Instituto de Astrofisica
de Canarias. This research is part of the Thesis of Degree of N. Gennari
(Bologna University). This research is partially supported by the italian
{Ministero  dell'Universit\'a e della Ricerca Scientifica} (MURST) through the
COFIN grant p.  2002028935-001, assigned to the project  {\em Distance and
stellar populations in the galaxies of the Local Group}. The finantial support
of the Agenzia Spaziale Italiana (ASI) is also acknowledged. Y. Momany is
acknowledged for providing his unpublished photometry of Leo~II that we used to
check our photometric calibration. We are grateful to an anonymous Referee for
his/her useful comments and indications.
Part of the data analysis has been performed
using software developed by P. Montegriffo at the INAF - Osservatorio
Astronomico di Bologna. This research has made use of data products from the
2MASS, a joint project of Univ. of Massachusetts, IPAC/CIT funded by NASA and
NSF. This research has made use of data products from the Guide Star
Catalogue,  produced at STScI.   This research has made use of NASA's
Astrophysics Data System Abstract Service.

\label{lastpage}

\end{document}